\documentclass[journal,11pt,onecolumn]{IEEEtran}
\usepackage{epsfig}
\usepackage{subfigure}
\usepackage{graphicx}
\usepackage{times}
\usepackage{cite}
\usepackage{enumerate}
\usepackage{amssymb}
\usepackage{algorithm}
\usepackage{algorithmic}
\usepackage{multirow}
\usepackage{amsmath}
\usepackage{braket}
\newtheorem{corollary}{Corollary}[section]
\newtheorem{theorem}{Theorem}[section] 
\newtheorem{lemma}{Lemma}[section]

\begin{document}

\title{Channel-Aware Random Access in the Presence of Channel Estimation Errors}

\author{
\authorblockN{Jeongho Jeon,~\IEEEmembership{Student Member,~IEEE,} and Anthony Ephremides,~\IEEEmembership{Life Fellow,~IEEE}}
\thanks{The material in this paper was presented in part at the IEEE International Symposium on Information Theory, Cambridge, MA, USA, July 2012.}
\thanks{The authors are with the
Department of Electrical and Computer Engineering and the Institute
for Systems Research, University of Maryland, College Park, MD 20742
USA (e-mail: jeongho@umd.edu; etony@umd.edu).}}
\maketitle

\begin{abstract}
In this work, we consider the random access of nodes adapting their transmission probability based on the local channel state information (CSI) in a decentralized manner, which is called CARA. The CSI is not directly available to each node but estimated with some errors in our scenario. Thus, the impact of imperfect CSI on the performance of CARA is our main concern. Specifically, an exact stability analysis is carried out when a pair of bursty sources are competing for a common receiver and, thereby, have interdependent services. The analysis also takes into account the compound effects of the multipacket reception (MPR) capability at the receiver. The contributions in this paper are twofold: first, we obtain the exact stability region of CARA in the presence of channel estimation errors; such an assessment is necessary as the errors in channel estimation are inevitable in the practical situation. Secondly, we compare the performance of CARA to that achieved by the class of stationary scheduling policies that make decisions in a centralized manner based on the CSI feedback. It is shown that the stability region of CARA is not necessarily a subset of that of centralized schedulers as the MPR capability improves.

\end{abstract}

\begin{keywords}
Channel-aware random access, channel estimation errors, stability region, multipacket reception capacity
\end{keywords}

\section{Introduction}

\PARstart{I}{ncreasing} demand for high data rate to support a wide range of services in wireless data networks has led to the exploitation of diversity amongst users \cite{knopp:information}. The diversity gain arises from the fact that wireless links experience random fading due to the constructive/destructive effect of multipath signal propagation and, thereby, there is always a user having better channel quality than the others at any time \cite{viswanath:opportunistic}. A downlink scheduler exploiting such diversity gain is called the opportunistic scheduler \cite{agrawal:optimality, refKusCon1, agrawal:class}.
%
%
Similar concept can be applied to the uplink communication but it is needed to have a centralized controller gathering channel state information (CSI) from distributed users, making a centralized decision, and distributing the decision information back to the distributed users.

The ALOHA protocol, the simple scheme of attempting transmission randomly, independently, distributively, and based on simple ACK/NACK feedback from the receiver, has gained continued popularity in distributed multiaccess communication systems since its creation \cite{abramson:aloha}. It is mainly due to its simplicity and the fact that it does not require centralized controllers. It also provided a basis for following enhanced schemes such as the Carrier Sense Multiple Access (CSMA) that combined ideas of fixed allocation with reservations via contention through random access. 
%
%
However, the random access systems have been built upon simplistic assumptions on the physical-layer such as the collision channel model, and the diversity in channel quality across different users was mostly out of consideration. Noticeably, there is a recent line of work on exploiting CSI under random access framework, which is called channel-aware random access (CARA) \cite{Qin06distributedapproaches, Adireddy02exploitingdecentralized, hong:stability, fanous:transmission}. The CARA allows the distributed nodes to adjust their random access probability based on the local CSI.

To put our contribution in perspective, we start with some background on CARA. In \cite{Qin06distributedapproaches}, it is assumed that each user has perfect local CSI and transmits only when the channel gain exceeds a certain threshold. The main contribution of the work is the characterization of the throughput scaling law for the system with infinitely backlogged users, i.e., users have packets to transmit at any time. Limitations of this work include the collision assumption made for the analysis when two or more users transmit at the same time. This assumption cannot be validated in general wireless communications since a transmission may succeed even in the presence of interference \cite{zorzi:capture, nguyen:capture, namislo:analysis, hajek:capture}. Also users were assumed to be symmetric in channel statistics. In \cite{Adireddy02exploitingdecentralized}, a similar problem with that in \cite{Qin06distributedapproaches} was considered but with additional multipacket reception (MPR) capability \cite{tong:multipacket, naware:stability, ghez:stability}. The advent of the multiuser detection technique for separating signals from the superposition of multiple received signals enables a receiver to correctly decode more than one packets simultaneously transmitted from different users. However, like \cite{Qin06distributedapproaches}, this work is based on the assumptions that users are always backlogged and symmetric in channel statistics. 
%
%
Note that in the system with bursty input traffic, it is not straightforward to know how users interfere with each other since they transmit only when having non-empty queues, which is more realistic than assuming always backlogged queues. In \cite{hong:stability}, the stability region\footnote{Stability region is a set of arrival rates maintaining the network queues in a bounded region. A formal definition is given in Section \ref{sec:system}.} of the system comprised of users having bursty packet arrivals and asymmetric channel statistics was obtained under the collision assumption and further extended in \cite{fanous:transmission} to the case with MPR capability. However, the analysis in both \cite{hong:stability} and \cite{fanous:transmission} is limited to the two-user scenario. This limitation is mainly due to the complex interaction between network queues. More specifically, the service process of individual queues depends on the status of the others. This is why most previous work on the stability of interacting systems has been focused on small-sized network and only bounds or approximations are known for networks with an arbitrary number of users \cite{tsybakov:ergodicity, rao:stability, luo:stability, szpankowski:stability, jeon:energy_mpr, jeon:cara, bordenave:asymptotic}.


In this work, we focus attention on the impact of imperfect CSI on the stability region of CARA for bursty input traffic. It needs to be emphasized that the analysis in all the above-mentioned work on CARA was performed based on the ideal assumption that the perfect local CSI is available at each user \cite{Qin06distributedapproaches, Adireddy02exploitingdecentralized, hong:stability, fanous:transmission}. In reality, however, the CSI is obtained through an estimation 
%
%
and any kind of estimation is imperfect as long as there is randomness in the observed signal. Consequently, the occurrence of errors in estimation is inevitable and the performance of CARA would certainly depend on the accuracy of channel estimation. To see the effect of imperfect CSI, we allow channel estimation errors in the two-state time-varying channel model considered in \cite{hong:stability, fanous:transmission}, in which each user $i$ transmits with probability $p_i$ when its channel state is \textit{good} and the packet queue is non-empty. Note that the errors either deprives the chance to utilize the \textit{good} channel state when falsely estimated to be \textit{bad} or causes unnecessary interference to the other user when its actual channel state is \textit{bad} but falsely estimated to be \textit{good}. There is some related work on CARA with imperfect CSI; in \cite{wang:stability}, the two-user system was considered under the collision channel model and, in \cite{wang:transmission}, a system with an arbitrary number of users was considered but with always backlogged queues and symmetric channel statistics. 

%

Our contribution in this work can be summarized as follows. We first introduce the realistic effect of practical channel estimation into the stability analysis of CARA. The analysis also takes into account the compound effect of the MPR capability, which depends not only on the set of transmitters but also on their instantaneous channel states. The derived stability region describes the theoretical limit on rates that can be pushed into the system while maintaining the queues stable at given channel estimation error rates and MPR probabilities. Secondly, by comparing with the case of having perfect CSI, we identify the loss due to the imperfect CSI on the stability region of CARA. Finally, the stability region of the \textit{longest connected queue} (LCQ) policy \cite{tassiulas:stability}, which is a \textit{throughput-optimal} policy that can stabilize the system whenever the stability is attainable, is derived again in the presence of channel estimation errors. The LCQ policy schedules a user having longest queue among those whose channel is \textit{connected} and, thus, requires queue length and channel state information feedback to the centralized controller. Interestingly, we observed that the stability region of CARA, a fully distributed policy, is not always a proper subset of that of the LCQ policy. This is when relatively strong MPR capability presents.

The rest of the paper is organized as follows. In Section \ref{sec:system}, we present the system model and revisit the notion of stability. In Section \ref{sec:main}, we describe our main result on the stability region of CARA in the presence of channel estimation errors. The proof of our main result is presented in Section \ref{sec:analysis} which is based on the \textit{stochastic dominance technique} previously introduced in \cite{rao:stability} to deal with interacting queues. In Section \ref{sec:schedulers}, the stability region of CARA is compared to that of the LCQ policy. Finally, we draw some conclusions in Section \ref{sec:conclusion}. 

\section{System Model}\label{sec:system}

\begin{figure}[t]
\centering
\epsfig{file=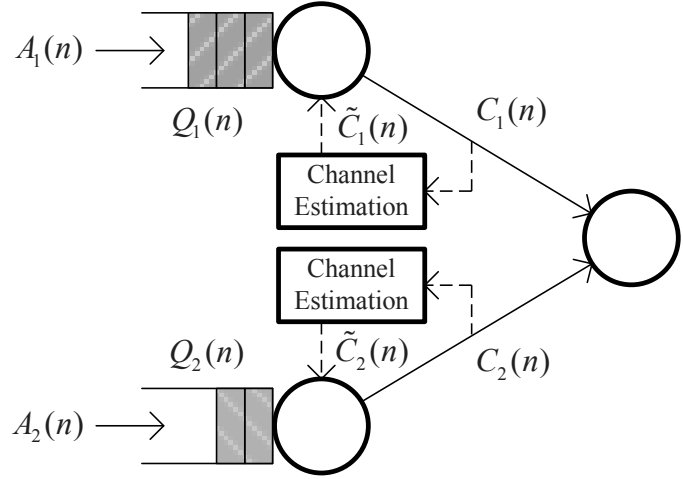,angle=0,width=0.5\textwidth}
\caption{Illustration of the system with bursty packet arrivals and channel estimation}
\label{fig:sys_model}
\vspace{0cm}
\end{figure}

We consider a multi-access system consisting of $N$ nodes and a common receiver. The stability analysis of CARA is done for $N=2$ as illustrated in Fig. \ref{fig:sys_model}, whereas $N$ is set to an arbitrary positive integer for the analysis of the LCQ policy in Section \ref{sec:schedulers}. Each node $i$ has an infinite size queue for storing the arriving packets that have fixed length. Time is slotted and the slot duration is equal to one packet transmission time. Let $Q_i(n)$ denote the number of packets buffered at $s_i$ at the beginning of the $n$-th slot which evolves according to
\begin{equation}\label{eqn:queueing_dynamics}
Q_i(n+1) = \max[Q_i(n) - \mu_i(n), 0] + A_i(n)
\end{equation}
where the stochastic processes $\{\mu_i(n)\}_{n=0}^{\infty}$ and $\{A_i(n)\}_{n=0}^{\infty}$ are sequences of random variables representing the number of arrivals and services at $s_i$ during time slot $n$, respectively. The arrival process $\{A_i(n)\}_{n=0}^{\infty}$ is modeled as an independent and identically distributed (i.i.d.) Bernoulli process with $E[A_i(n)]=\lambda_i$, and the processes at different nodes are assumed to be independent of each other. The service process $\{\mu_i(n)\}_{n=0}^{\infty}$ depends jointly on the transmission protocol and the underlying channel model, which governs the success of transmissions.

The channel between node $i$ and the receiver is randomly time-varying and its state at time slot $n$ is denoted by $C_i(n)$ and let $\boldsymbol{C}(n) = \{C_i(n), \dots, C_N(n) \}$. We assume that channels hold their state for the duration of a slot and potentially change on the slot boundaries\footnote{This assumption can be validated when the channel coherence time is relatively longer than the slot duration.}. As in the previous work \cite{hong:stability, fanous:transmission}, we model the time-varying channel as a discrete-time stochastic process taking values from $\{G,B\}$, which represents \textit{good} and \textit{bad} states, respectively. The channel processes at different nodes are assumed to be independent of each other but the realizations of a channel at a node at different time slots are not necessarily independent. If fact, a channel process can be arbitrarily correlated over time as long as stationary. We denote by $\pi_i^G$ and $\pi_i^B (= 1 - \pi_i^G)$ the steady-state probabilities that channel $i$ is in \textit{good} and \textit{bad} states, respectively. The transmission control policy considered in \cite{fanous:transmission} is studied again in which node $i$ transmits with probability $p_i$ when $C_i(n) = G$ and its queue is non-empty. We denoted by $\tilde{C}_i(n)$ the estimated channel state over the link between node $i$ and the receiver, and node $i$ is now transmitting with probability $p_i$ if $\tilde{C}_i(n) = G$ and its queue is non-empty. It is obvious that the performance of this adaptation would highly depend on the accuracy of the channel estimation. Let us define $\epsilon_i^G = \mathrm{Pr} [\tilde{C}_i(n) = B | C_i(n) = G ]$ and $\epsilon_i^B = \mathrm{Pr} [\tilde{C}_i(n) = G | C_i(n) = B ]$,
%
which are the probabilities of falsely estimating the channel state and let $\bar{\epsilon}_i^G = 1 - \epsilon_i^G$ and $\bar{\epsilon}_i^B = 1 - \epsilon_i^B$.

The success of a transmission depends on the underlying channel model. The MPR model used in this work enables the probabilistic reception of simultaneously transmitted packets owing to the multi-user detection \cite{verdu:multiuser}. Unlike the static MPR model used in \cite{tong:multipacket, naware:stability, ghez:stability}, in which the time-varying nature of wireless channels was disregarded, our model captures not only the effect of interference but also the instantaneous channel states of the transmitters. Denote by $\mathcal{N}_{\mathrm{tx}}$ and $\boldsymbol{C}_{\mathrm{tx}}(n)$ the set of transmitting nodes and their channel states, respectively. Then, the probability that a packet transmitted from node $i$ is successfully decoded at the destination is described by \begin{equation*}
q_{i|\boldsymbol{C}_{\mathrm{tx}}(n)} = \mathrm{Pr}[\gamma_{i|\boldsymbol{C}_{\mathrm{tx}}(n)} > \theta]
\end{equation*}
where $\gamma_{i|\boldsymbol{C}_{\mathrm{tx}}(n)}$ is the signal-to-interference-plus-noise-ratio (SINR) of the signal transmitted from node $i$ at the receiver given the channel states of the transmitters $\boldsymbol{C}_{\mathrm{tx}}(n)$ and the threshold for the successful decoding $\theta$, which depends on the modulation scheme, target bit-error-rate, and the number of bits in the packet \cite{goldsmith:wireless}. We assume throughout the paper that the success probability when a node's own channel state is \textit{bad} is negligible due to such as the deep fading regardless of the interference. This is the situation when the signal-to-noise-ratio (SNR) itself is below the threshold $\theta$ for the successful decoding. Thus, for $N=2$, we are particularly interested in the following reception probabilities\footnote{For example, $q_{2|\{B,G\}}$ is the probability that the transmission by node 2 is successful when $C_1(n)=B$ and $C_2(n)=G$.}
\begin{equation*}
q_{1|\{G\}}, q_{1|\{G,B\}}, q_{1|\{G,G\}}, q_{2|\{G\}}, q_{2|\{B,G\}}, q_{2|\{G,G\}}
\end{equation*}
Note that the success probability when a signal is transmitted in the presence of interference cannot exceed the probability when it is transmitted alone. Moreover, since we are considering the multi-access to a common receiver, the fact that a node's channel is in the \textit{good} state implies potentially higher interference level to the other node if they transmit at the same time. Therefore, the following relations hold:
\begin{math}
q_{1|\{G\}} > q_{1|\{G,B\}} > q_{1|\{G,G\}}
\end{math}
and
\begin{math}
q_{2|\{G\}} > q_{2|\{B,G\}} > q_{2|\{G,G\}}.
\end{math}

We adopt the notion of stability used in \cite{szpankowski:stability} in which the stability of a queue is equivalent to the existence of a proper limiting distribution. In other words, a queue is said to be \textit{stable} if
\begin{equation*}\label{eqn:definition_stability}
    \lim_{n \rightarrow \infty} \mathrm{Pr}[Q_i(n) < {x} ] = F(x) \ \ \mathrm{and} \ \
    \lim_{ {x} \rightarrow \infty} F(x) = 1
\end{equation*}
%
%
Loynes' theorem, as it relates to stability, plays a central role in our approach \cite{loynes:stability}. It states that if the arrival and service processes of a queue are strictly jointly
stationary and the average arrival rate is less than the average
service rate, the queue is stable. 
%
%
For given channel estimation error rates $\boldsymbol{\epsilon}$ and transmission probabilities $\boldsymbol{p}$ of the nodes, the stability region $\mathfrak{S}(\boldsymbol{\epsilon}, \boldsymbol{p})$ is defined as the set of arrival rate vectors $\boldsymbol{\lambda}=(\lambda_1, \lambda_2)$ for which all queues in the system are stable. The stability region of the system $\mathfrak{S}(\boldsymbol{\epsilon})$ is defined as the closure of $\mathfrak{S}(\boldsymbol{\epsilon}, \boldsymbol{p})$ over all possible transmission probability vectors, i.e., $\mathfrak{S}(\boldsymbol{\epsilon}) \triangleq \bigcup_{\boldsymbol{p} \in [0,1]^2}
    \mathfrak{S}(\boldsymbol{\epsilon}, \boldsymbol{p})$.

\section{CARA with Imperfect CSI}\label{sec:main}

This section describes the stability region of CARA in the presence of channel estimation errors. As noted earlier, the service process of a queue depends on the status of the other, which makes the analysis challenging. The proof of the results, which is based on the stochastic dominance technique \cite{rao:stability}, is presented in the next section.
%
%
Let us define
\begin{align*}
\Psi_1^{\boldsymbol{\epsilon}} & = \pi_2^{G}\bar{\epsilon}_2^G \left( q_{1|\{G\}} - q_{1|\{G,G\}} \right)  + \pi_2^{B} \epsilon_2^B \left( q_{1|\{G\}} - q_{1|\{G,B\}}\right)  \\ 
\Psi_2^{\boldsymbol{\epsilon}} & = \pi_1^{G} \bar{\epsilon}_1^G \left( q_{2|\{G\}} - q_{2|\{G,G\}} \right)  + \pi_1^{B} \epsilon_1^B \left( q_{2|\{G\}} - q_{2|\{B,G\}}\right) 
\end{align*}
which are shorthand notations to simplify the description of our main results.

\begin{lemma}\label{lem:stability_region}
The stability region $\mathfrak{S}(\boldsymbol{\epsilon}, \boldsymbol{p})$ of CARA at given channel estimation error rate vector $\boldsymbol{\epsilon}$ and transmission probability vector $\boldsymbol{p}$ is the union of the following subregions:
%
\begin{equation*}\label{eqn:sr_fixed_r1}
    \mathcal{R}_1=\left\{ (\lambda_{1},\lambda_{2}): \lambda_1 < \pi_{1}^{ G} \bar{\epsilon}_1^G p_1 \left( q_{1|\{G\}}  - \frac{\Psi_1^{\boldsymbol{\epsilon}} \lambda_2 }{ \pi_{2}^{G} \bar{\epsilon}_2^G \left( q_{2|\{G\}} - \Psi_2^{\boldsymbol{\epsilon}} p_1 \right) }  \right), \lambda_2 < \pi_2^{G} \bar{\epsilon}_2^G p_2 \left( q_{2|\{G\}} -\Psi_2^{\boldsymbol{\epsilon}} p_1 \right) \right\}
\end{equation*}
and
\begin{equation*}\label{eqn:sr_fixed_r2}
    \mathcal{R}_2=\left\{ (\lambda_{1},\lambda_{2}): \lambda_1 < \pi_1^{G} \bar{\epsilon}_1^G p_1 \left( q_{1|\{G\}} -\Psi_1^{\boldsymbol{\epsilon}} p_2 \right), \lambda_2 <  \pi_{2}^{ G} \bar{\epsilon}_2^G p_2 \left( q_{2|\{G\}}  - \frac{\Psi_2^{\boldsymbol{\epsilon}} \lambda_1 }{ \pi_{1}^{G} \bar{\epsilon}_1^G \left( q_{1|\{G\}} - \Psi_1^{\boldsymbol{\epsilon}} p_2 \right) }  \right) \right\}
\end{equation*}

\begin{proof}
The proof is presented in the next section.
\end{proof}
\end{lemma}

Let us define the following points in the two-dimensional Euclidean space: 
\begin{align}
P_{1} & = \left( \frac{ \pi_1^{G} \bar{\epsilon}_1^G q_{2|\{G\}}(q_{1|\{G\}} - \Psi_1^{\boldsymbol{\epsilon}})^2}{\Psi_2^{\boldsymbol{\epsilon}} q_{1|\{G\}}}, \frac{\pi_2^{G} \bar{\epsilon}_2^G \Psi_1^{\boldsymbol{\epsilon}} q_{2|\{G\}}}{q_{1|\{G\}}}\right) \label{eqn:P1}\\
P_{2} & = \left( \frac{\pi_1^{G}\bar{\epsilon}_1^G \Psi_2^{\boldsymbol{\epsilon}} q_{1|\{G\}}}{q_{2|\{G\}}}, \frac{ \pi_2^{G} \bar{\epsilon}_2^G q_{1|\{G\}}(q_{2|\{G\}} - \Psi_2^{\boldsymbol{\epsilon}})^2}{\Psi_1^{\boldsymbol{\epsilon}} q_{2|\{G\}}} \right) \label{eqn:P2}\\
P_{3} & = (\pi_1^{G}\bar{\epsilon}_1^G (q_{1|\{G\}} - \Psi_1^{\boldsymbol{\epsilon}}), \pi_2^{G} \bar{\epsilon}_2^G (q_{2|\{G\}} - \Psi_2^{\boldsymbol{\epsilon}})) \label{eqn:P3}
\end{align}
which are all in the first quadrant. 

\begin{theorem}\label{thm:random_access_w_errors}
If $\frac{\Psi_1^{\boldsymbol{\epsilon}}}{q_{1|\{G\}}} + \frac{\Psi_2^{\boldsymbol{\epsilon}}}{q_{2|\{G\}}}  \geq 1$, the boundary of the stability region $\mathfrak{S}(\boldsymbol{\epsilon})$ of CARA at a given channel estimation error rate vector $\boldsymbol{\epsilon}$ is described by three segments: (i) the line connecting $P_Y = (0, \pi_2^{G} \bar{\epsilon}_2^G q_{2|\{G\}})$ and $P_{1}$, (ii) the curve
\begin{equation}\label{eqn:stability_curve_1}
\sqrt{\frac{\Psi_2^{\boldsymbol{\epsilon}}}{\pi_1^{G} \bar{\epsilon}_1^G} \lambda_1} + \sqrt{\frac{\Psi_1^{\boldsymbol{\epsilon}}}{\pi_2^{G}\bar{\epsilon}_2^G } \lambda_2}  = \sqrt{q_{1|\{G\}} q_{2|\{G\}}}
\end{equation}
from $P_{1}$ to $P_{2}$, and (iii) the line connecting $P_{2}$ and $P_X = (\pi_1^{G}\bar{\epsilon}_1^G q_{1|\{G\}}, 0)$. If $\frac{\Psi_1^{\boldsymbol{\epsilon}}}{q_{1|\{G\}}} + \frac{\Psi_2^{\boldsymbol{\epsilon}}}{q_{2|\{G\}}}  < 1$, it is described by two lines: (i) the line connecting $P_Y$ and $P_{3}$ and (ii) the line connecting $P_{3}$ and $P_X$. 

\begin{proof}
The proof is presented in the next section.
\end{proof}
\end{theorem}
%

%
\begin{corollary}
If $\frac{\Psi_1^{\boldsymbol{\epsilon}}}{q_{1|\{G\}}} + \frac{\Psi_2^{\boldsymbol{\epsilon}}}{q_{2|\{G\}}} >1$, the stability region $\mathfrak{S}(\boldsymbol{\epsilon})$ is non-convex. If $\frac{\Psi_1^{\boldsymbol{\epsilon}}}{q_{1|\{G\}}} + \frac{\Psi_2^{\boldsymbol{\epsilon}}}{q_{2|\{G\}}} \leq 1$, it is a convex polygon. Specifically, when $\frac{\Psi_1^{\boldsymbol{\epsilon}}}{q_{1|\{G\}}} + \frac{\Psi_2^{\boldsymbol{\epsilon}}}{q_{2|\{G\}}} = 1$, the region becomes a right triangle.
\end{corollary}
This corollary can be easily verified by comparing the slopes of the lines from $P_Y$ to $P_{1}$ and from $P_{2}$ to $P_X$ and those from $P_Y$ to $P_{3}$ and from $P_{3}$ to $P_X$. Specifically, if $\frac{\Psi_1^{\boldsymbol{\epsilon}}}{q_{1|\{G\}}} + \frac{\Psi_2^{\boldsymbol{\epsilon}}}{q_{2|\{G\}}} = 1$, the curve \eqref{eqn:stability_curve_1} shrinks to a point whose coordinate is identically described by both $P_{1}$ and $P_{2}$ and the slopes of the lines from $P_Y$ to $P_{1}$ and from $P_{2}$ to $P_X$ become identical.

Consider the case when perfect CSI is available. This can be viewed as a special case of our model with $\boldsymbol{\epsilon} = \boldsymbol{0}$, where $\boldsymbol{0}$ is the vector of zeros. By substituting $\boldsymbol{\epsilon} = \boldsymbol{0}$ into Theorem \ref{thm:random_access_w_errors}, we can obtain the stability region for the case with perfect CSI, which reconfirms the previous results obtained in \cite{hong:stability} and \cite{fanous:transmission}. For the comparison's sake, let us consider the case when CSI is not available and, hence, each node has to make decisions on transmission independent of the underlying channel states. This corresponds to the original ALOHA in which each node transmits with probability $p_i$ regardless of the underlying channel states whenever its queue is non-empty. Thus, at a given set of transmitters, the success probability of each node is given as a constant, which is obtained by taking the average over the stationary distribution of the channel states. Denote by $q_i^{\mathrm{s}}$ and $q_i^{\mathrm{m}}$ $(i \in \set{1,2})$ the transmission success probabilities seen by node $i$ when it transmits alone or along with the other node $j(\neq i)$. For the two-node case, it is obtained as ${q}_1^{\mathrm{s}} = \pi_{1}^G q_{1|\{G\}}$, ${q}_1^{\mathrm{m}} = \pi_{1}^G \pi_{2}^G q_{1|\{G,G\}} + \pi_{1}^G \pi_{2}^B  q_{1|\{G,B\}}$, ${q}_2^{\mathrm{s}} = \pi_{2}^G q_{2|\{G\}}$,
and ${q}_2^{\mathrm{m}} = \pi_1^G \pi_{2}^G q_{2|\{G,G\}} + \pi_{1}^B \pi_{2}^G q_{2|\{B,G\}}$. Also define $\Delta_ i  = q_i^{\mathrm{s}} - q_i^{\mathrm{m}}$, which is assumed to be strictly positive without loss of generality. The following theorem obtained in \cite{naware:stability} describes the stability region of the original ALOHA for the case with static MPR channels, but it is also applicable to the system with time-varying channels but when the CSI is unavailable. This is because the success probabilities are given as constants over time. We especially denote the stability region for this case by $\mathfrak{S}(\emptyset)$ since the notion of channel estimation errors for the case with no CSI is not valid.


%
\begin{figure}\label{fig:sr_case1}
\centering
\epsfig{file=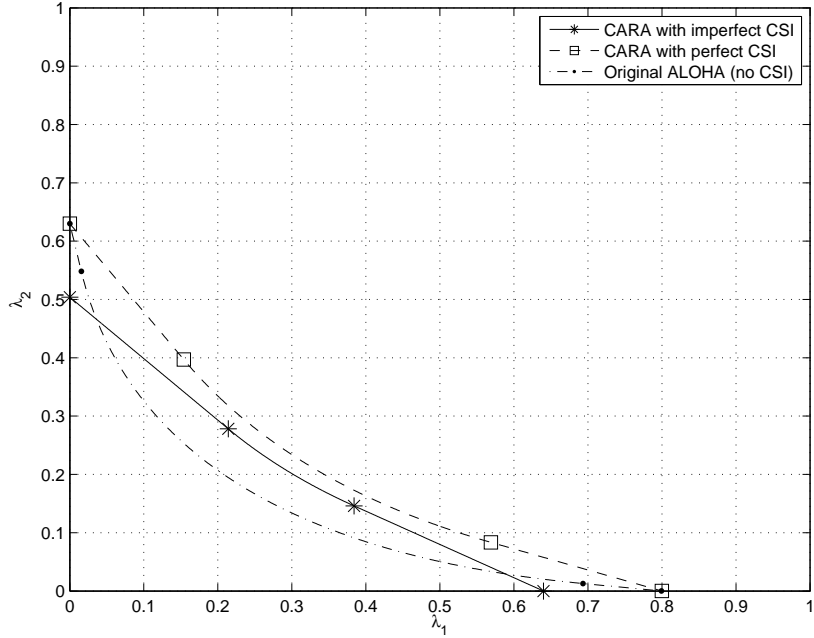,angle=0,width=0.7\textwidth}
\caption{Stability region of CARA: the case of non-convex region (parameter setting:$\pi_{1}^G = 0.8$, $\pi_{2}^G = 0.7$, $q_{1|\{G\}} = 1$, $q_{2|\{G\}} = 0.9$, $q_{1|\{G,B\}} =q_{2|\{B,G\}}= 0.2$, $q_{1|\{G,G\}} = q_{2|\{G,G\}} = 0.1$, $\epsilon_i^j = 0.2, \forall i,j$)}
\end{figure}
\begin{figure}\label{fig:sr_case2}
\centering
\epsfig{file=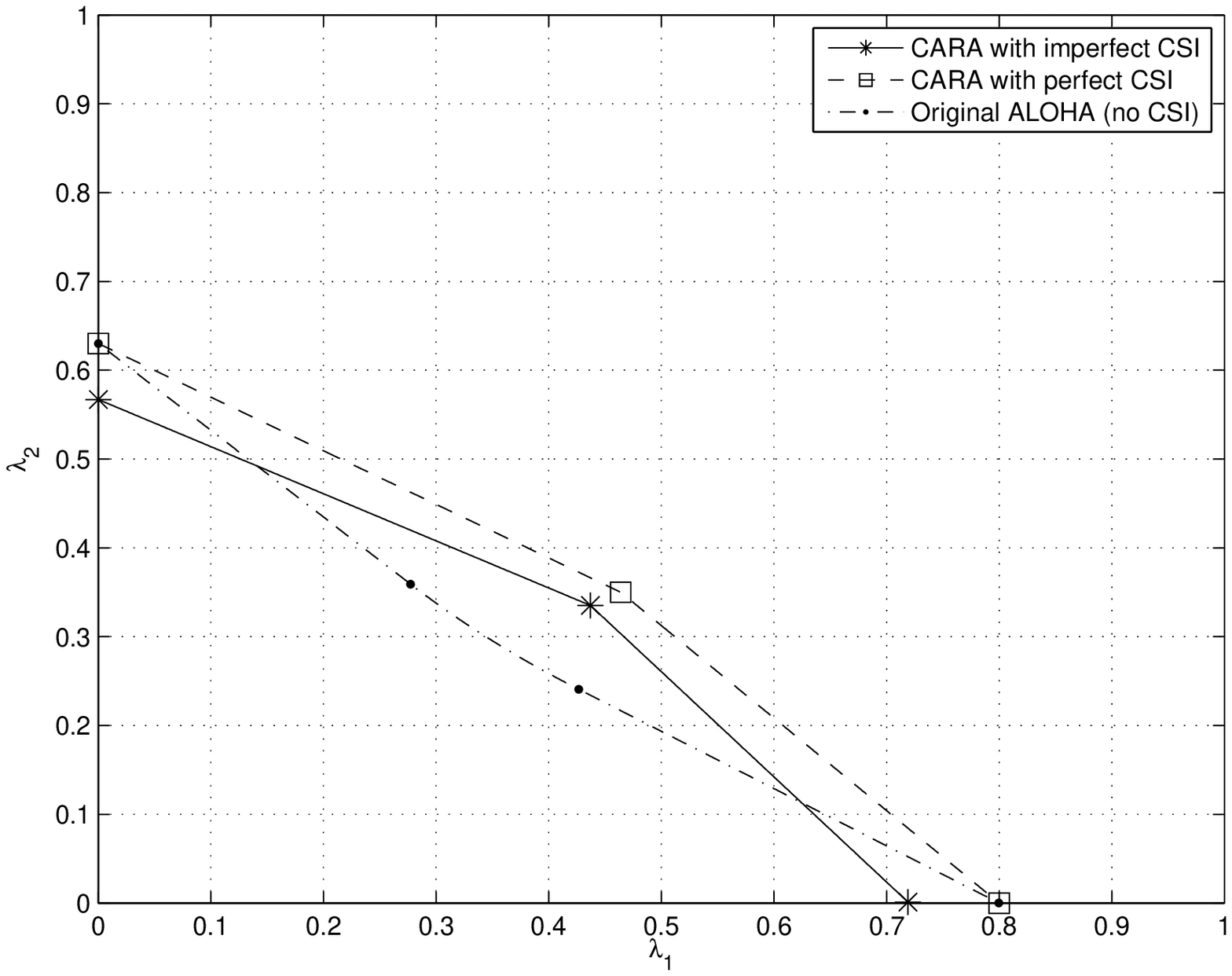,angle=0,width=0.7\textwidth}
\caption{Stability region of CARA: the case of convex region (parameter setting:$\pi_{1}^G = 0.8$, $\pi_{2}^G = 0.7$, $q_{1|\{G\}} = 1$, $q_{2|\{G\}} = 0.9$, $q_{1|\{G,B\}} =q_{2|\{B,G\}}= 0.5$, $q_{1|\{G,G\}} = q_{2|\{G,G\}} = 0.4$, $\epsilon_i^j = 0.1, \forall i,j$)}
\end{figure}
\begin{theorem}\label{thm:original_aloha}
If $\frac{\Delta_1}{q_1^{\mathrm{s}}} + \frac{\Delta_2}{q_2^{\mathrm{s}}}  \geq 1$, the boundary of the stability region of ALOHA with no CSI, denoted by $\mathfrak{S}(\emptyset)$, is described by three segments: (i) the line connecting $P_Y = (0, q_2^{\mathrm{s}})$ and $P_{1} = (\frac{q_2^{\mathrm{s}} (q_1^{\mathrm{s}} - \Delta_1)^2}{\Delta_2 q_1^{\mathrm{s}}}, \frac{\Delta_1 q_2^{\mathrm{s}}}{q_1^{\mathrm{s}}})$, (ii) the curve
\begin{math}\label{eqn:stability_curve}
\sqrt{\Delta_2 \lambda_1} + \sqrt{\Delta_1 \lambda_2} = \sqrt{q_1^{\mathrm{s}} q_2^{\mathrm{s}}}
\end{math}
from $P_{1}$ to $P_{2} = (\frac{\Delta_2 q_1^{\mathrm{s}}}{q_2^{\mathrm{s}}}, \frac{q_1^{\mathrm{s}}(q_2^{\mathrm{s}} -\Delta_2)^2}{\Delta_1 q_2^{\mathrm{s}}})$, and (iii) the line connecting $P_{2}$ and $P_X = (q_1^{\mathrm{s}}, 0)$. If $\frac{\Delta_1}{q_1^{\mathrm{s}}} + \frac{\Delta_2}{q_2^{\mathrm{s}}} < 1$, it is described by two lines: (i) the line connecting $P_Y$ and $P_{3} = (q_1^{\mathrm{s}} - \Delta_1, q_2^{\mathrm{s}} - \Delta_2)$ and (ii) the line connecting $P_{3}$ and $P_X$. 

\begin{proof}
Please refer to \cite{naware:stability}. 
\end{proof}
\end{theorem}

In Fig. 1 and Fig. 2, we illustrate the stability region of CARA with imperfect CSI along with the case with perfect CSI and the original ALOHA with no CSI. It is evident that the stability region of CARA with perfect CSI always includes that of original ALOHA, which complies with the previous results obtained in \cite{hong:stability}. It is also obvious that the stability region of CARA with perfect CSI always includes that with imperfect CSI, and the difference between the two regions, therefore, can be understood as the loss due to the errors in channel estimation. However, the stability region of original ALOHA is not a proper subset of that of CARA with imperfect CSI as shown in the figures. The inefficiency of CARA is due to the fact that each node transmits only when the channel is estimated to be \textit{good} and those time slots when the channel is estimated to be \textit{bad} but when it is indeed \textit{good} is not exploited.

\section{Analysis Using Stochastic Dominance}\label{sec:analysis}

In this section, we provide details on the derivation of our main result presented in the previous section. Note that nodes are interfering with each other only when they are transmitting and, under the considered protocol, node $i$ transmits with probability $p_i$ if the estimated channel state is \textit{good} and its queue is non-empty. Consequently, the service process of a queue at a node depends on the status of the queue at the other node and, thus, the content size of the queues form a two-dimensional discrete-time Markov chain with infinite state space, which makes the problem challenging even for the two-node case. Having to deal with a system with such \textit{interacting} queues, a tool that can be used is the stochastic dominance technique which was introduced in \cite{rao:stability} and further used in \cite{naware:stability, szpankowski:stability, jeon:energy_mpr, jeon:cara}. The essence of the stochastic dominance technique is to decouple the interaction between queues via the construction of a hypothetical system; this hypothetical system operates as follows: i) the packet arrivals at each node occur at \textit{exactly} the same instants as in the original system, ii) the \textit{coin tosses} that determine the transmission attempts of the nodes have \textit{exactly} the same outcomes in both systems, iii) however, one of the nodes in the system continues to transmit dummy packets even when its packet queue is empty. Sending dummy packets is only aimed to cause constant interference to the other node and does not contribute to throughput if the transmission is successful. It is obvious that sample-pathwise the queue sizes in this dominant system will never be smaller than their counterparts in the original system, provided the queues start with identical initial conditions. Thus, the stability condition obtained for the dominant system is a sufficient condition for the stability of the original system. It turns out, however, that it is indeed sufficient and necessary, which will be discussed in detail later in this section.


Construct a hypothetical system which is identical to the original system except that node $2$ transmits dummy packets when it decides to transmit but its packet queue is empty. Define $1_{i}(n)$ to be an indicator function whose value is one if the transmission by node $i$ is successful, which also necessarily requires that the corresponding node transmits at that time slot. Otherwise, $1_{i}(n)=0$. By conditioning on the underlying actual channel states, the average service rate of the queue at node $1$ can be expressed as
\begin{align}\label{eqn:d12_service_1}
\mu_1 & = \sum_{\boldsymbol{C}'} \mathrm{Pr} [1_{1} | \boldsymbol{C} = \boldsymbol{C}'] \mathrm{Pr}[\boldsymbol{C} = \boldsymbol{C}'] \nonumber \\
& = \mathrm{Pr} [1_{1} | \boldsymbol{C} = \{G,G\} ] \pi_1^G \pi_2^G + \mathrm{Pr} [1_{1} | \boldsymbol{C} = \{G, B\} ] \pi_1^G \pi_2^B
\end{align}
where we used the fact that the transmission success probability is zero when a node's own channel state is \textit{bad} and the time index $n$ is suppressed in the steady-state. Note that each node $i$ transmits with probability $p_i$ only when the estimated channel state is \textit{good}. By further conditioning on the estimated channel states, it is expressed as
\begin{equation*}
\mathrm{Pr} [1_{1} | \boldsymbol{C} = \boldsymbol{C}' ] = \sum_{\tilde{\boldsymbol{C}}'} \mathrm{Pr} [1_{1} | \boldsymbol{C} = \boldsymbol{C}', \tilde{\boldsymbol{C}} = \tilde{\boldsymbol{C}}' ] \mathrm{Pr}[\tilde{\boldsymbol{C}} = \tilde{\boldsymbol{C}}' | \boldsymbol{C} = \boldsymbol{C}']
\end{equation*}
where
\begin{equation}\label{eqn:conditional_success_1} 
\mathrm{Pr} [1_{1} | \boldsymbol{C} = \{G,G\} ] = \left[ q_{1|\{G\}} p_1 (1 - p_2) + q_{1|\{G,G\}} p_1 p_2\right] \bar{\epsilon}_1^G \bar{\epsilon}_2^G + q_{1|\{G\}} p_1 \bar{\epsilon}_1^G {\epsilon}_2^G
\end{equation}
and
\begin{equation}\label{eqn:conditional_success_2}
\mathrm{Pr} [1_{1} | \boldsymbol{C} = \{G, B\} ] = q_{1|\{G\}} p_1 \bar{\epsilon}_{1}^G \bar{\epsilon}_{2}^B + \left[ q_{1|\{G\}} p_1 (1 - p_2) + q_{1|\{B,G\}} p_1 p_2\right] \bar{\epsilon}_{1}^G {\epsilon}_{2}^B 
\end{equation}
%
%
%
By substituting \eqref{eqn:conditional_success_1} and \eqref{eqn:conditional_success_2} into \eqref{eqn:d12_service_1} and, after some manipulations, the average service rate of the queue at node 1 is derived as
\begin{equation*}\label{eqn:d12_service_rate2}
\mu_1 = \pi_1^{G} \bar{\epsilon}_1^G p_1 \left( q_{1|\{G\}} -\Psi_1^{\boldsymbol{\epsilon}} p_2 \right)
\end{equation*}
where $\Psi_i^{\boldsymbol{\epsilon}}$ is defined in Section \ref{sec:main}. By Loynes' Theorem, the queue at node 1 is stable if $\lambda_1 < \mu_1$. Note that the queue size at node 1 in this dominant system can be modeled as a discrete-time $M$/$M$/1 system with arrival rate $\lambda_1$ and the service rate given above, which does not depend on the status of the queue at node 2. For stable input rate $\lambda_1$ that is less than $\mu_1$, the queue at node 1 empties out with probability 
\begin{equation*}
\mathrm{Pr}[Q_1 = 0] = 1 - \frac{\lambda_1}{\mu_1} = 1 - \frac{\lambda_1}{\pi_1^{G} \bar{\epsilon}_1^G p_1 \left( q_{1|\{G\}} -\Psi_1^{\boldsymbol{\epsilon}} p_2 \right)}
\end{equation*}
Observe that the service process of the queue at node 2 depends on the status of the queue at node 1 since node 1 is able to transmit only when its queue is non-empty and, thereby, interfering with node 2. By conditioning on the emptiness of the queue at node 1, the average service rate of the queue at node 2 can be expressed as
\begin{equation}\label{eqn:d1_expression_service_rate1}
\mu_2 = \mathrm{Pr} [1_{2} | Q_1 \neq 0] \mathrm{Pr}[Q_1 \neq 0] + \mathrm{Pr} [1_{2} | Q_1 = 0] \mathrm{Pr}[Q_1 = 0]
\end{equation}
The service rate of the queue at node 2 when the queue at node 1 is non-empty can be obtained by following the same procedure used for deriving the service rate of the queue at node 1 and is given by
\begin{equation}\label{eqn:d1_expression_service_rate1_t1}
\mathrm{Pr} [1_{1} | Q_2 \neq 0] = \pi_1^{G}\bar{\epsilon}_1^G p_1 \left( q_{1|\{G\}} -\Psi_2^{\boldsymbol{\epsilon}} p_1 \right)
\end{equation}
The service rate of the queue at node 2 when the queue at node 1 is empty can be obtained quite simply as it does not depend on the action made by node 1 and is given by
\begin{equation}\label{eqn:d1_expression_service_rate1_t2}
\mathrm{Pr} [1_{1} | Q_2 = 0] = \pi_1^{G} \bar{\epsilon}_1^G p_1 q_{1|\{G\}}
\end{equation}
Substituting \eqref{eqn:d1_expression_service_rate1_t1} and \eqref{eqn:d1_expression_service_rate1_t2} into \eqref{eqn:d1_expression_service_rate1} yields the average service rate of the queue at node 2 which can be summarized to
\begin{equation*}\label{eqn:d12_service_2}
\mu_2 =  \pi_{2}^{ G} \bar{\epsilon}_2^G p_2 \left( q_{2|\{G\}}  - \frac{\Psi_2^{\boldsymbol{\epsilon}} \lambda_1 }{ \pi_{1}^{G}\bar{\epsilon}_1^G \left( q_{1|\{G\}} - \Psi_1^{\boldsymbol{\epsilon}} p_2 \right) }  \right)
\end{equation*}
and by Loyne's Theorem, the queue at node 2 is stable if $\lambda_2 < \mu_2$. Consequently, the stable input rate pairs $(\lambda_1, \lambda_2)$ are those less than $(\mu_1, \mu_2)$ elementwise, and it gives the description of $\mathcal{R}_2$ in Lemma \ref{lem:stability_region}. By reversing the roles of the two nodes, we construct a parallel dominant system in which node 1 transmits dummy packets and, by following the same procedure used for the first dominant system, the stability region for this parallel dominant system is obtained as described in Lemma \ref{lem:stability_region}, which is denoted by $\mathcal{R}_1$.

Importantly, the stability condition obtained using the dominant system technique is not merely a sufficient condition for the stability of the original system but is sufficient and necessary. Consider, for example, the subregion $\mathcal{R}_2$ obtained for the dominant system in which node 2 transmits dummy packets. The sufficient part is trivial which follows from the construction of the dominant system such that the dominant system stochastically dominates the original system in number of packets in the queues at all times. The necessary part can be proved as follows: if for some $\lambda_2$, the queue at node 2 is unstable in the hypothetical system, then $Q_2(n)$ approaches infinity almost surely. Note that as long as the queue does not empty, the behavior of the hypothetical system and the original system are identical, provided they start from the same initial conditions, since dummy packets will never have to be used. A sample-path that goes to infinity without visiting the empty state, which is a feasible one for a queue that is unstable, will be identical for both the hypothetical and the original systems. Therefore, the instability of the hypothetical system implies the instability of the original system.

We now obtain the maximum achievable stability region, which is the closure of the stability region $\mathfrak{S}(\boldsymbol{\epsilon}, \boldsymbol{p})$ over the transmission probability vector $\boldsymbol{p}$. An equivalent way of taking the closure operation is to optimize the boundary of the stability region $\mathfrak{S}(\boldsymbol{\epsilon}, \boldsymbol{p})$ over $\boldsymbol{p}$. For the subregion $\mathcal{R}_2$, for example, we set up the following optimization problem, in which $\mu_2$ is maximized over $\boldsymbol{p}$ at given $\lambda_1$ while satisfying the stability of the queue at node 1.
\begin{align}
\max_{\boldsymbol{p}} \ & \ \mu_2 = \pi_{2}^{ G} \bar{\epsilon}_2^G p_2 \left( q_{2|\{G\}}  - \frac{\Psi_2^{\boldsymbol{\epsilon}} \lambda_1 }{ \pi_{1}^{G}  \bar{\epsilon}_1^G \left( q_{1|\{G\}} - \Psi_1^{\boldsymbol{\epsilon}} p_2 \right) }  \right) \label{eqn:opt_obj}\\
\displaystyle \textrm{subject to} \ & \  \lambda_1 < \pi_1^{G}  \bar{\epsilon}_1^G p_1 \left( q_{1|\{G\}} -\Psi_1^{\boldsymbol{\epsilon}} p_2 \right) \label{eqn:opt_const_1}\\
\displaystyle \ & \vec{p} \in [0, 1]^2 \label{eqn:opt_const_2}
\end{align}
Note that $\mu_2$ depends only on $p_2$ but not on $p_1$. Differentiating $\mu_2$ with respect to $p_2$ gives
\begin{equation*}\label{eqn:y_first_derivative}
\frac{\partial \mu_2}{\partial p_2} = \pi_2^{G}\bar{\epsilon}_2^G \left( q_{2|\{G\}} - \frac{\Psi_2^{\boldsymbol{\epsilon}} q_{1|\{G\}} \lambda_1}{\pi_{1}^{G} \bar{\epsilon}_1^G \left( q_{1|\{G\}} - \Psi_1^{\boldsymbol{\epsilon}} p_2 \right)^2} \right)
\end{equation*}
By differentiating once again, we have
\begin{equation*}
\frac{\partial^2 \mu_2}{\partial p_2^2} = - \frac{ 2 \Psi_1^{\boldsymbol{\epsilon}} \Psi_2^{\boldsymbol{\epsilon}} \pi_2^G \bar{\epsilon}_2^G q_{1|\{G\}} \lambda_1 }{\pi_1^G \bar{\epsilon}_1^G \left( q_{1|\{G\}} - \Psi_1^{\boldsymbol{\epsilon}} p_2 \right)^3} 
\end{equation*}
Observe that
\begin{align}\label{eqn:inequalites}
q_{1|\{G\}} - \Psi_1^{\boldsymbol{\epsilon}} p_2 & \geq q_{1|\{G\}} - \Psi_1^{\boldsymbol{\epsilon}} \nonumber \\
& = (1 - \pi_2^G \bar{\epsilon}_2^G - \pi_2^B {\epsilon}_2^B) q_{1|\{G\}}  + \pi_2^{G} \bar{\epsilon}_2^G q_{1|\{G,G\}} + \pi_2^{B} \epsilon_2^B q_{1|\{G,B\}} \nonumber \\
& \geq (1 - \pi_2^G  - \pi_2^B ) q_{1|\{G\}}  + \pi_2^{G} \bar{\epsilon}_2^G q_{1|\{G,G\}} + \pi_2^{B} \epsilon_2^B q_{1|\{G,B\}} \nonumber \\
& = \pi_2^{G} \bar{\epsilon}_2^G q_{1|\{G,G\}} + \pi_2^{B} \epsilon_2^B q_{1|\{G,B\}} \nonumber \\
& > 0
\end{align}
Therefore, the second derivative is strictly negative and, consequently, $\mu_2$ is a concave function of $p_2$. Equating the first derivative to zero gives the maximizing $p_2^{\ast}$ as
\begin{equation}\label{eqn:optimal_p2}
p_2^{\ast} = \frac{1}{\Psi_1^{\boldsymbol{\epsilon}}} \left( q_{1|\{G\}} - \sqrt{\frac{\Psi_2^{\boldsymbol{\epsilon}} q_{1|\{G\}} \lambda_1}{\pi_1^{G} \bar{\epsilon}_1^G q_{2|\{G\}}}} \right)
\end{equation}
and the corresponding maximum function value is obtained by substituting \eqref{eqn:optimal_p2} into \eqref{eqn:opt_obj} as 
\begin{equation*}\label{eqn:max_y_curve}
\mu_{2,\mathrm{curve}}^{\ast} = \frac{\pi_2^{G}\bar{\epsilon}_2^G}{\Psi_1^{\boldsymbol{\epsilon}}} \left( q_{1|\{G\}} - \sqrt{\frac{\Psi_2^{\boldsymbol{\epsilon}} q_{1|\{G\}} \lambda_1}{\pi_1^{G} \bar{\epsilon}_1^G q_{2|\{G\}}}} \right) \left( q_{2|\{G\}} - \sqrt{\frac{\Psi_2^{\boldsymbol{\epsilon}} q_{2|\{G\}} \lambda_1}{\pi_1^{G} \bar{\epsilon}_1^G q_{1|\{G\}}}} \right) 
\end{equation*}
Suppose that the maximum occurs at a strictly interior point of the feasible region, i.e., $p_2^{\ast} \in (0,1)$, which corresponds to the condition
\begin{equation}\label{eqn:feasible_range_1}
\frac{\pi_1^{G} \bar{\epsilon}_1^G q_{2|\{G\}}(q_{1|\{G\}} - \Psi_1^{\boldsymbol{\epsilon}})^2}{\Psi_2^{\boldsymbol{\epsilon}} q_{1|\{G\}}} < \lambda_1 < \frac{\pi_1^{G} \bar{\epsilon}_1^G q_{1|\{G\}} q_{2|\{G\}}}{\Psi_2^{\boldsymbol{\epsilon}}}
\end{equation}
that is obtained by rearranging \eqref{eqn:optimal_p2} and substituting the extreme values of $p_2^{\ast}$, i.e., 0 and 1. On the other hand, the constraint \eqref{eqn:opt_const_1} should also be satisfied for the derived $p_2^{\ast}$, which gives
\begin{equation}\label{eqn:feasible_range_2}
\lambda_1 < \frac{\pi_1^{G}\bar{\epsilon}_1^G \Psi_2^{\boldsymbol{\epsilon}} q_{1|\{G\}}}{q_{2|\{G\}}}
\end{equation}
Consequently, $\mu_{2,\mathrm{curve}}^{\ast}$ is valid only for the intersection of \eqref{eqn:feasible_range_1} and \eqref{eqn:feasible_range_2}. By comparing the endpoints, the intersection is specified to be the same with \eqref{eqn:feasible_range_1} if $\Psi_2^{\boldsymbol{\epsilon}} \geq q_{2|\{G\}}$, which is impossible. This is because, from the relation described in \eqref{eqn:inequalites}, it can be deduced that $\Psi_i^{\boldsymbol{\epsilon}} < q_{i|\{G\}}$. If $\frac{\Psi_1^{\boldsymbol{\epsilon}}}{q_{1|\{G\}}} + \frac{\Psi_2^{\boldsymbol{\epsilon}}}{q_{2|\{G\}}} \geq 1$, the intersection becomes
\begin{equation*}
\frac{\pi_1^{G} \bar{\epsilon}_1^G q_{2|\{G\}}(q_{1|\{G\}} - \Psi_1^{\boldsymbol{\epsilon}})^2}{\Psi_2^{\boldsymbol{\epsilon}} q_{1|\{G\}}}  < \lambda_1 < \frac{\pi_1^{G} \bar{\epsilon}_1^G \Psi_2^{\boldsymbol{\epsilon}} q_{1|\{G\}}}{q_{2|\{G\}}}
\end{equation*}
Otherwise, it is an empty set. 

Next suppose that $p_2^{\ast}$ is either $0$ or $1$, which is the case when $\lambda_1$ is outside of the range \eqref{eqn:feasible_range_1}. If $\lambda_1$ is on the right-hand side of the range, $\frac{\partial \mu_2}{\partial p_2}$ becomes non-positive and, thus, $\mu_2$ is a non-increasing function of $p_2$. Therefore, $p_2^{\ast} = 0$, which gives $\mu_2^{\ast} = 0$. Whereas, if $\lambda_1$ is on the left-hand side of the range, $\mu_2$ is a non-decreasing function of $p_2$ and, thus, $p_2^{\ast} = 1$. The corresponding maximum function value is obtained as
\begin{equation*}
\mu_{2,\mathrm{line}}^{\ast} = \pi_2^{G} \bar{\epsilon}_2^G \left( q_{2|\{G\}} - \frac{\Psi_2^{\boldsymbol{\epsilon}} \lambda_1}{\pi_1^{G} \bar{\epsilon}_1^G (q_{1|\{G\}} -\Psi_1^{\boldsymbol{\epsilon}})} \right)
\end{equation*}
%
%
By substituting $p_2^{\ast} = 1$ into \eqref{eqn:opt_const_1}, we have 
\begin{equation}\label{eqn:feasible_range_4}
\lambda_1 < \pi_1^{G} \bar{\epsilon}_1^G (q_{1|\{G\}} - \Psi_1^{\boldsymbol{\epsilon}})
\end{equation} 
Given the fact that $\lambda_1$ lies on the left-hand side of the range \eqref{eqn:feasible_range_1} in addition to the above constraint, it is shown that if $\frac{\Psi_1^{\boldsymbol{\epsilon}}}{q_{1|\{G\}}} + \frac{\Psi_2^{\boldsymbol{\epsilon}}}{q_{2|\{G\}}} \geq 1$, $\mu_{2,\mathrm{line}}^{\ast} $ is valid for $\lambda_1$ on the entire range of the left-hand side of \eqref{eqn:feasible_range_1}. Otherwise, it is valid for the range specified by \eqref{eqn:feasible_range_4}. Following the similar procedure, we can optimize the boundary of the subregion $\mathcal{R}_1$, which completes the proof of Theorem \ref{thm:random_access_w_errors}.
%


\section{Comparisons with the Centralized Schedulers}\label{sec:schedulers}

In this section, we compare the stability region of CARA to that achieved by scheduling policies that make centralized decision based on the CSI feedback. Note that a scheduler allocates each time slot to one of the nodes such that the scheduled node can transmit in an interference-free environment during the allocated slot. In \cite{tassiulas:dynamic}, it was shown that queue length information can be utilized to improve the scheduling performance. Specifically, the discovered policy that serves the \textit{longest-connected-queue} (LCQ) among those in `\textit{Good}' channel state stabilizes the system whenever the input rate vector is inside the stability region. Here, the stability region is defined as the set of arrival rate vectors that can be stably supported by considering all possible stationary scheduling policies. This is why the LCQ policy is called a \textit{throughput-optimal} policy. The following theorem is derived again for the case with channel estimation errors, which was originally derived in \cite{tassiulas:dynamic} for the case with perfect CSI.

\begin{theorem}\label{thm:lcq}
The necessary and sufficient stability condition by considering all possible stationary scheduling policies in the presence of channel estimation errors is 
\begin{equation}\label{eqn:stability_lcq}
\sum_{i \in \mathcal{N}'} \frac{\lambda_i}{q_{i|\{G\}}} < 1 - \prod_{i \in \mathcal{N}'} \left( 1 - \pi_i^G \bar{\epsilon}_i^G\right), \ \ \ \forall \mathcal{N}' \in \set{1,\dots, N}
\end{equation}
Furthermore, the LCQ policy stabilizes the system as long as it is stabilizable.

\begin{proof}
Assume that the system is operating under certain stationary policy and it is stable. Denote by $1_i(n)$ the indicator function that is equal to 1 if the transmission by node $i$ is successful, which necessarily implies that node $i$ is chosen to transmit at that time slot. Also denote by $I_i(n)$ the indicator function that is equal to 1 if the actual channel state between node $i$ and the receiver is \textit{Good} and it is estimated correctly. The expectations are given by $E[1_i(n)] = q_{i|\{C_i(n)\}}$, and $E[I_i(n)] = \pi_i^G \bar{\epsilon}_i^G$. The number of packets at queue $i$ evolves with time according to the queueing dynamics in equation \eqref{eqn:queueing_dynamics} with $\mu_i(n) = h_i(n) \cdot 1_i(n)$,
where $h_i(t) = 1$ if node $i$ is scheduled at time slot $n$ and $I_i(n) = 1$. Thus, the departure rate from queue $i$ is written as
\begin{equation*}
E[h_i(t) 1_i(t)] = q_{i|\{G\}} E[h_i(t)]
\end{equation*}
Note that if the system is stable, the rate of what comes in must be equal to the rate of what goes out. In other words, for any subset of network nodes $\mathcal{N}'$, the following equality must hold.
\begin{equation}\label{eqn:lcq_app_temp1}
\sum_{i \in \mathcal{N}'} \frac{\lambda_i}{q_{i|\{G\}}} = \sum_{i \in \mathcal{N}'} E[h_i(t)]
\end{equation}
Consider now the partition of the probability space into the events
\begin{align*}
\Omega_1 & = \set{I_i(t) = 0, i \in \mathcal{N}'} \\
\Omega_2 & = \set{I_i(t) = 0, i \in \mathcal{N}'}^c \cap \set{Q_i(t-1) = 0, i \in \mathcal{N}'} \\
\Omega_3 & = \set{I_i(t) = 0, i \in \mathcal{N}'}^c \cap \set{Q_i(t-1) = 0, i \in \mathcal{N}'}^c
\end{align*}
where $\Omega^c$ is the complementary set of $\Omega$. Notice that $E\left[ \sum_{i \in \mathcal{N}'} h_i(t) | \Omega_l \right] = 0$ for $l = 1,2$ and, therefore, we have
\begin{align*}
\sum_{i \in \mathcal{N}'} E[h_i(t)] & = E \left[ E\left[ \sum_{i \in \mathcal{N}'} h_i(t) | \Omega_3 \right] \mathrm{Pr}[\Omega_3] \right] \\
& < 1 - \mathrm{Pr}[\Omega_1] - \mathrm{Pr}[\Omega_2]
\end{align*}
Owing to the assumption on the independence of the channel processes between different nodes, we have
\begin{equation*}
\mathrm{Pr}[\Omega_1] = \prod_{i \in \mathcal{N}'} (1 - \pi_i^G \bar{\epsilon}_i^G)
\end{equation*}
and $\mathrm{Pr}[\Omega_2] > 0$. Therefore, we have
\begin{equation}\label{eqn:lcq_app_temp2}
\sum_{i \in \mathcal{N}'} E[h_i(t)]  < 1 - \prod_{i \in \mathcal{N}'} (1 - \pi_i^G \bar{\epsilon}_i^G)
\end{equation}
and equations \eqref{eqn:lcq_app_temp1} and \eqref{eqn:lcq_app_temp2} implies equation \eqref{eqn:stability_lcq} in the theorem.

For the sufficiency, we use the Lyapunov drift argument similar to that used in \cite{tassiulas:dynamic}. Denote by $L(\vec{Q}(t))$ a Lyapunov function of the queue length process and define the conditional Lyapunov drift as $\Delta(L(\vec{Q}(t))) = E [L(\vec{Q}(t+1)) - L(\vec{Q}(t)) | \vec{Q}(t)]$. If there exist some $\epsilon >0$ and a finite number $b$ such that the conditional Lyapunov drift satisfies $\Delta(L(\vec{Q}(t))) < - \epsilon$ for $L(\vec{Q}(t))>b$, the queues in the system are stable. This is an application of Foster's criterion for ergodicity of a Markov chain \cite{tassiulas:stability}. Let us consider $L(\vec{Q}) = \sum_{i=1}^N q_{i|\{G\}}^{-1} Q_i^2$. Then, for the considered Lyapunov function, it can be shown that the conditional Lyapunov drift satisfies
\begin{equation}\label{eqn:lcq_app}
\Delta(L(\vec{Q}(t))) < 1 + \sum_{i=1}^N q_{i|\{G\}}^{-1} \left( E[A_i^2(t)] + 2 \lambda_i Q_i(t)\right) - 2 E \left[ \sum_{i=1}^N Q_i(t) h_i(t+1) | \vec{Q}(t) \right]
\end{equation}
The details on the derivation of the above inequality can be found in \cite{tassiulas:dynamic}.

Define now a permutation $e_i$ for $i = 0, \dots, N$ such that $e_0 = 0$ and $Q_{e_i} (t) \geq Q_{e_{i-1}}(t)$, for $i = 2, \dots, N$. Consider also a partition of the probability space into the events $\Phi_i$ defined by
\begin{equation*}
\Phi_ i = \set{I_{e_i}(t+1) = 1, I_{e_j}(t+1) = 0, \ N\geq j >i}
\end{equation*}
for $i  = 1, \dots, N$ and $\Phi_0 = \set{ \vec{I}(t+1) = \vec{0}}$. Notice that from the definition of the LCQ policy, in the event $\Phi_j$, queue $e_j$ is served if it is not empty. Therefore, the last term in the right-hand side of \eqref{eqn:lcq_app} becomes
\begin{align*}
E \left[ \sum_{i=1}^N Q_i(t) h_i(t+1) | \vec{Q}(t) \right] = \sum_{i=1}^N Q_{e_i}(t) \pi_i^G \bar{\epsilon}_i^G \prod_{j = i+1}^N (1 - \pi_{e_j}^G \bar{\epsilon}_{e_j}^G)
\end{align*}
By substituting the above into \eqref{eqn:lcq_app} and after some manipulation, we obtain
\begin{equation*}\label{eqn:drift_app}
\Delta(L(\vec{Q}(t))) < 1 + \sum_{i=1}^N E[A_i^2(t)] + 2 Q_{e_N}(t) \max_{\mathcal{N}' \subset \set{1, \dots, N}}  \left[ \sum_{i \in \mathcal{N}'} \frac{\lambda_i}{q_{i|\{G\}}} -1 + \prod_{i \in \mathcal{N}'} \left( 1 - \pi_i^G \bar{\epsilon}_i^G\right) \right]
\end{equation*}
It is not difficult to observe that if the condition \eqref{eqn:stability_lcq} is met, then for sufficiently large $Q_{e_N}(t)$, the right-hand side of the above conditional Lyapunov drift becomes negative. This completes the proof.
\end{proof}
\end{theorem}
\begin{figure}\label{fig:lcqd}
\centering
\epsfig{file=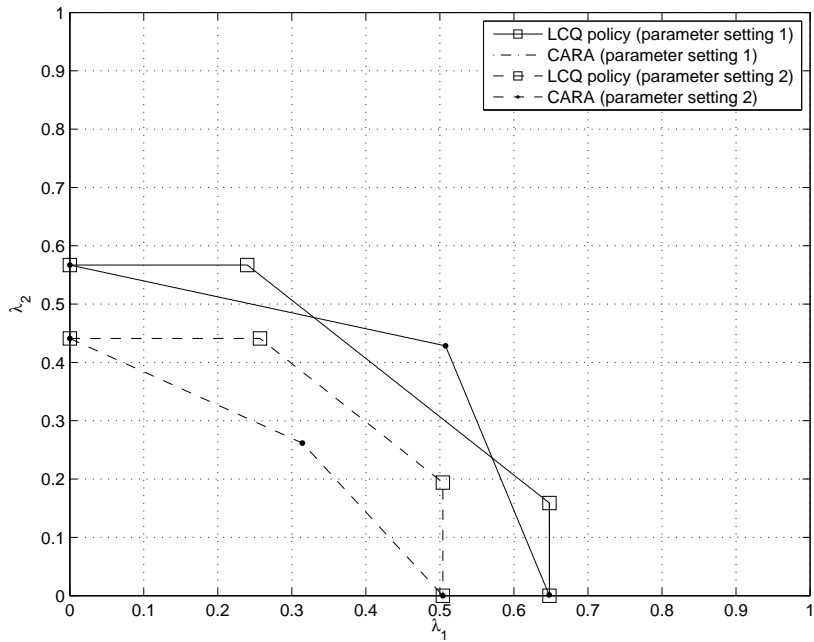,angle=0,width=0.7\textwidth}
\caption{Comparison with the LCQ policy (common setting: $\pi_{1}^G = 0.8$, $\pi_{2}^G = 0.7$, $q_{1|\{G\}} = q_{2|\{G\}} = 0.9$, parameter setting 1: $q_{1|\{G,B\}} =q_{2|\{B,G\}}= 0.7$, $q_{1|\{G,G\}} = q_{2|\{G,G\}} = 0.6$, $\epsilon_i^j = 0.1, \forall i,j$, parameter setting 2: $q_{1|\{G,B\}} =q_{2|\{B,G\}}= 0.4$, $q_{1|\{G,G\}} = q_{2|\{G,G\}} = 0.3$, $\epsilon_i^j = 0.3, \forall i,j$)}
\end{figure}

For the two-node case, the stability condition \eqref{eqn:stability_lcq} in the above theorem becomes
\begin{equation*}
\frac{\lambda_1}{q_{1|\{G\}}} + \frac{\lambda_2}{q_{2|\{G\}}}< \pi_1^G \bar{\epsilon}_1^G + \pi_2^G \bar{\epsilon}_2^G(1- \pi_1^G \bar{\epsilon}_1^G) 
\end{equation*}
and ${\lambda_i} < \pi_i^G \bar{\epsilon}_i^G {q_{i|\{G\}} }$ for $i \in \{1,2\}$. In Fig. 2, we compare the stability region of CARA to that achieved by LCQ policy, and it can be observed that the former is not necessarily a subset of the latter. Indeed, the relationship between them depends on parameters such as the channel estimation error and the MPR probabilities. Note that the stability region of CARA becomes a subset of that of the LCQ policy when the stability region of CARA is non-convex, or convex but if the vertex $P_3$ given in \eqref{eqn:P3} is strictly contained in the stability region achieved by LCQ policy. The condition for $P_3$ to be inside the stability region of the LCQ policy is given by 
\begin{equation*}\label{eqn:condition_subset}
\frac{\Psi_1^{\boldsymbol{\epsilon}} }{\pi_2^G \bar{\epsilon}_2^G q_{1|\{G\}}} + \frac{\Psi_2^{\boldsymbol{\epsilon}}}{\pi_1^G \bar{\epsilon}_1^G  q_{2|\{G\}}} > 1
\end{equation*}
Otherwise, if the stability region of CARA is convex and the above inequality does not hold, the stability region of CARA is not a proper subset of that of the LCQ policy, i.e., there exists a region that can be achieved only by CARA.

\section{Concluding Remarks}\label{sec:conclusion}

In this work, we studied the stability property of CARA in the presence of channel estimation errors and showed that its stability region may not strictly contain that of original ALOHA. To guarantee the superiority of CARA even with imperfect CSI, we need to modify the protocol itself such that each node transmits with some positive probability, although it believes that the channel is in the \textit{bad} state. Such modification was not considered here. We also compared the stability region of CARA to that achieved by the throughput-optimal LCQ policy and showed that the former is not necessarily a subset of the latter especially as the MPR capability improves. The stability analysis of CARA had to resolve the complex interaction between nodes and, hence, extending the results to the general case with an arbitrary number of nodes, although highly desirable, presents serious difficulties and was not considered here.

\bibliographystyle{IEEEtran}
\bibliography{bib_all}

\end{document}